\documentclass{sig-alternate-2013}
\newfont{\mycrnotice}{ptmr8t at 7pt}
\newfont{\myconfname}{ptmri8t at 7pt}
\let\crnotice\mycrnotice%
\let\confname\myconfname%

\permission{Permission to make digital or hard copies of all or part of this work for personal or classroom use is granted without fee provided that copies are not made or distributed for profit or commercial advantage and that copies bear this notice and the full citation on the first page. Copyrights for components of this work owned by others than ACM must be honored. Abstracting with credit is permitted. To copy otherwise, or republish, to post on servers or to redistribute to lists, requires prior specific permission and/or a fee. Request permissions from Permissions@acm.org.}
\conferenceinfo{ICTIR '15,}{September 27-30, 2015, Northampton, MA, USA}
\copyrightetc{\copyright~2015 ACM. \the\acmcopyr}
\crdata{ISBN 978-1-4503-3833-2/15/09\ ...\$15.00.\\
DOI: http://dx.doi.org/10.1145/2808194.2809458}

\usepackage{color}
\usepackage{colortbl}
\usepackage{pgf}
\usepackage{tikz}
\usepackage{pgfplots}
\usepackage{footnote}
\usepackage{url}
\usepackage{multirow}
\newcommand{\sub}{\textbf{s}}

\newcommand{\obj}{\textbf{o}}

\newcommand{\miss}{\textbf{--}}

\begin{document}

\title{Entropy and Graph Based Modelling of Document Coherence using Discourse Entities: An Application to IR}
%
%
\numberofauthors{2} 
%
\author{
%
%
\alignauthor
Casper Petersen\\
       \affaddr{Department of Computer Science}\\
       \affaddr{University of Copenhagen, Denmark}\\
       \email{cazz@di.ku.dk}
\alignauthor
Christina Lioma\\
       \affaddr{Department of Computer Science}\\
       \affaddr{University of Copenhagen, Denmark}\\
       \email{c.lioma@di.ku.dk}
\and
\alignauthor
Jakob Grue Simonsen\\
       \affaddr{Department of Computer Science}\\
       \affaddr{University of Copenhagen, Denmark}\\
       \email{simonsen@di.ku.dk}
\alignauthor
Birger Larsen\\
       \affaddr{Department of Communication}\\
       \affaddr{Aalborg University Copenhagen, Denmark}\\
       \email{birger@hum.aau.dk}
}
\date{30 July 1999}
%
\maketitle
\begin{abstract}
We present two novel models of document coherence and their application to information retrieval (IR). Both models approximate document coherence using discourse 
entities, e.g. the subject or object of a sentence.
Our first model views text as a Markov process generating 
sequences of discourse entities (entity $n$-grams); we use the entropy of these 
entity $n$-grams to approximate the rate at which new 
information appears in text, reasoning that as more new words appear, the topic increasingly drifts and text
 coherence decreases. 
Our second model extends the work of Guinaudeau \& Strube~\shortcite{Guinaudeau:Strube:2013} that represents text as a graph of discourse entities, 
linked by different relations, such as their distance or adjacency in text. 
We use several graph topology metrics to approximate different aspects 
of the discourse flow that can indicate
coherence, such as the average clustering or betweenness of discourse entities in text.
Experiments with several instantiations of these models show that: (i) our models perform on a par with 
two other well-known models of text coherence even without any parameter tuning, and (ii) reranking retrieval results according to their coherence scores gives notable performance 
gains, confirming a relation between document coherence and relevance.
This work contributes two novel models of document coherence, the application of which to IR complements recent work in the integration of 
document cohesiveness or comprehensibility to ranking \cite{BenderskyCD11,TanGP12}. 
\end{abstract}

%

%
\section{Introduction}
\label{s:intro}
The extent to which text makes sense by introducing, explaining and linking its concepts and ideas through a sequence of semantically 
and logically related units of discourse is called \emph{coherence}. Several models of text coherence exist. 
On a high level, these models capture text regularities or patterns predicted by a theory or otherwise
hypothesised to indicate coherence. 
For instance, according to early discourse theories \cite{grosz:1986}, three core factors of text coherence are (i) the discourse purpose (\textit{intentional
structure}), (ii) the specific discourse items discussed (\textit{attentional structure}), and (iii) the organisation of the \textit{discourse
segments}. Of these factors, coherence models have been presented for both intentional structure \cite{LouisN12}, and discourse segments
\cite{BarzilayL04,FungN06}, but attentional structure has received the most attention and is the basis for most existing models of text coherence. One
approach to 
attentional structure that has been used extensively in coherence models is \emph{Centering Theory}~\cite{grosz:1983,joshi:1981} (CT), which
posits that a reader's attention is centered on a few salient entities in text, and that these exhibit patterns signalling the reader to switch or
retain attention. A widely used application of CT is the \emph{entity grid} model~\cite{Barzilay:2008}, which 
assumes that coherence can be measured from sequences of repeated discourse entities, such as the subject and object of a sentence. 
In this work, we use the entity grid as our basis to propose two novel classes of models
for document coherence.

\begin{figure}[]
\centering
\scalebox{0.28}{
\includegraphics[scale=0.6]{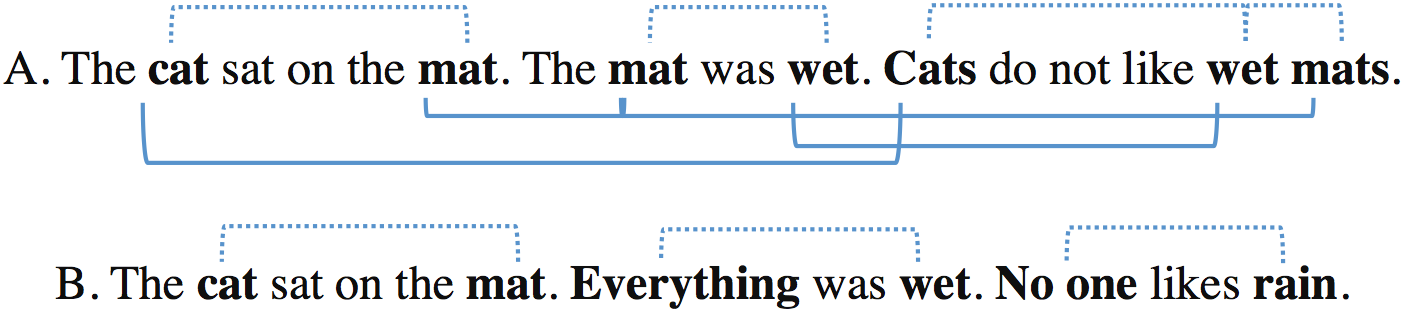}
}
\caption{Example of more (A) and less (B) coherent text. Discourse entities are in bold. Dotted lines mark \textit{within-sentence relations} 
of discourse entities. 
Solid lines mark \textit{between-sentence relations}, which newly introduced entities do not have, decreasing coherence.}
\label{fig:cat}
\end{figure}

Our first class of models calculates coherence using the information \textit{\textbf{entropy}}~\cite{shannon:2001} of the salient discourse entities 
in the
entity grid. The information entropy rate of a document can be thought of as the rate at which new information appears as one reads the document. The main 
idea is that, as more new information appears, the document becomes less focussed on a single topic, so its coherence decreases. 
This assumption, which is already verified in the literature \cite{BenderskyCD11}, is illustrated in Figure \ref{fig:cat}, which juxtaposes the discourse flow between three entities (\texttt{cat(s), mat(s), wet}) 
and six entities (\texttt{cat, mat, everything, wet, no one, rain}) in texts A and B respectively. As entities are linked between sentences, i.e. repeated, 
text becomes more topically focussed and hence more coherent. 
However, when many new entities are introduced, since they are by definition not linked to previous discourse, the topic of the text becomes less clear and overall coherence decreases.
  To
the best of our knowledge, discourse entropy has not been used for text coherence estimation before (there is however work on \textit{lexical entropy} for text 
readability, which we discuss in Section \ref{sec:discoursetheories}). 

Our second class of document coherence models maps the entity grid to \textit{\textbf{graphs}} following 
Guinaudeau \& Strube~\shortcite{Guinaudeau:Strube:2013}, 
where entities are vertices, linked 
by various relations between them, e.g. distance or adjacency. The topologies of these graphs model the flow of discourse.
We investigate whether graph properties, such as the clustering coefficient, or iterative graph ranking algorithms, such as 
PageRank, can approximate
document coherence. This class of models extends work by Guinaudeau \& Strube that used only a single such metric, namely outdegree, to calculate text coherence.

We present several instantiations of the above two classes of coherence models and we evaluate their effectiveness, first as coherence models per se, 
and second when integrated to information retrieval (IR). 
In the first case, we evaluate our coherence models in the standard \emph{sentence
reordering} task. We find that our models are more accurate than two well known baselines in coherence modelling (one of them being the entity grid~\cite{Barzilay:2008}
 that we extend), even without any 
parameter tuning or training. 
In the second case, we investigate whether factoring coherence in the IR process improves retrieval precision, 
on the basis that coherence should be a reasonable predictor of relevance. Experiments with standard TREC Web track data 
show that 
reranking retrieval results according to their coherence scores 
improves retrieval precision, especially for the top 20 retrieved results.	

In the rest of the paper, Section~\ref{sec:discoursetheories} overviews related work; 
Sections~\ref{sec:ecohmodel} \&~\ref{sec:graphcomodel} present our entropy and graph based coherence models respectively; 
Section~\ref{s:eval} discusses the experimental evaluation of these models, and Section \ref{s:discussion} their limitations and future extensions.
 Sections~\ref{s:conc} summarises our conclusions.

\section{Related Work}~\label{sec:discoursetheories}
We overview the Natural Language Processing (NLP) literature on models of text coherence (Section \ref{ss:localglobal}), focussing in particular on
 those using discourse entities 
(Section \ref{ss:disc}). We also discuss related work on applying text coherence to improve NLP tasks and IR (Section \ref{ss:quality}).

\subsection{Local and global coherence}
\label{ss:localglobal}
A text can be coherent at a \textit{local} and \textit{global} level~\cite{enos:1996}. \textit{Local} coherence is measured by examining the
similarity between neighbouring text spans, e.g., the well-connectedness of adjacent sentences through lexical cohesion \cite{HallidayH76}, or entity
repetition \cite{grosz:1995}. \textit{Global} coherence, on the other hand, is measured through discourse-level relations connecting remote text spans across a whole
text, e.g. sentences \cite{Kehler02,MannT98}.

There is extensive work on local coherence that uses different approaches, including bag of words methods at sentence level \cite{FoltzKL98}, 
sequences of content words (of length $\geq 3$) at paragraph level \cite{shin:2000}, 
 local lexical cohesion information \cite{BarzilayEM02}, local syntactic cues \cite{ElsnerC08a}, and combining local lexical and syntactic features, 
 e.g., term co-occurrence \cite{Lapata03,SoricutM06a}. Overall, various aspects of CT have long been used to model local coherence
\cite{KibbleP04,PoesioSEH04}, including the well-known entity approaches that rank the repetition and syntactic realisation of entities in adjacent
sentences \cite{Barzilay:2008,ElsnerC08a}.

There is also work on \textit{global} coherence, focussing on the structure of a document as a whole
\cite{BarzilayL04,ChenBBK09,ElsnerC08a,FungN06}. 
However, not many coherence models represent both local and global coherence, even though those two are connected: local coherence is a prerequisite
for global coherence~\cite{Barzilay:2008}, and there is psychological evidence that coherence on both local and global levels is manifested in text
comprehension \cite{Tapiero07,zhang:2011}. Among the few models that capture both local and global text coherence is the sentence ordering model of
Zhang \shortcite{zhang:2011}: on a local level, sentences are represented as  concept vectors where concepts are equivalent to content words; on a
global level, sentences are represented as vertices, and sentence relations as edges in a document graph. 

Our work
captures both local and global coherence as explained in Sections \ref{sec:ecohmodel} \& \ref{sec:graphcomodel}, which practically means that 
it can model document coherence both in the simple case of capturing adjacent discourse transitions like those illustrated in Figure \ref{fig:cat}, but also in 
the more complex (yet more realistic) case of capturing non-adjacent discourse transitions like those illustrated in Figure \ref{fig:natfmetrics}.

\subsection{Entity based coherence models} 
\label{ss:disc}
The basis of our two coherence models is the entity grid \cite{Barzilay:2008}, which 
represents a document by its salient discourse entities and their 
syntactic roles, see Table \ref{tbl:examplegrid} for example.
The entity grid indicates the location of each discourse entity in a document, which is important for coherence modelling because mentions of
an entity tend to appear in clusters of neighbouring sentences in coherent documents. This last assumption is adapted from CT where
consecutive utterances are regarded as more coherent if they keep mentioning the same entities~\cite{grosz:1995,PoesioSEH04}.
There are several extensions and variations of the entity grid, including adaptations for German \cite{FilippovaS07}, coupled with extensions
integrating high level sentence structure \cite{CheungP10}; extensions integrating writing quality features (e.g. word variety and style indicators)
\cite{burstein:2010}; extensions of the original entity grid (which captures sentence-to-sentence entity transitions) to capture term occurrences in
sentence-to-sentence relation sequences \cite{LinNK11}; extensions focussing on syntactic regularities \cite{LouisN12}; 
a \textit{topical entity grid} that considers topical information instead of only discourse entities \cite{ElsnerC11}; and extensions incorporating
modifiers and named entity types into the entity grid to distinguish important from less important entities \cite{Elsner:Charniak:2011}. A recent
extension \cite{Guinaudeau:Strube:2013} maps the entity grid into a graph where coherence is calculated as the average outdegree of the nodes in the
graph. 
While this model is only presented as a local coherence model, it does represent both local and global coherence: entities are represented as nodes in
a graph representing the document, allowing to connect entities occurring in non-adjacent sentences, hence spanning globally in the document. The
coherence models we present in this paper employ the
entity grid of \cite{Barzilay:2008}, but use several novel computations based on this grid: 
first entropy, and second an extension of the work of Guinaudeau \& Strube~\shortcite{Guinaudeau:Strube:2013} with several more complex graph 
metrics than outdegree.

\subsection{Applications of text coherence}
\label{ss:quality}
Apart from being an interesting problem in itself, coherence models and hence coherence prediction are important to a variety of NLP
 tasks and applications, such as summarisation \cite{BarzilayEM02,BarzilayL04,CelikyilmazH11,LinLNK12,zhang:2011}, machine translation
\cite{LinLNK12,xiong:2013}, 
 separating conversational threads \cite{ElsnerC11}, text fluency/reading difficulty detection~\cite{bicknell:2009,MiltsakakiK04,PitlerN08}, grammars
for natural 
 language generation \cite{Banik09,Kibble01,KibbleP04}, genre classification \cite{Barzilay:2008}, sentence insertion \cite{ChenSB07} and 
 sentence ordering \cite{Barzilay:2008,karamanis:2006,KaramanisMPO09}. In IR in particular, the document coherence scores produced by our 
 models can be seen 
as similar to several document quality measures that have been used in the past to improve retrieval. 
Examples of document quality include, for instance, document complexity, which Mikk \cite{Mikk:2001} 
predicts using a corrected term frequency measure based on word commonness. 
In the context of web search in particular, document quality has been studied extensively. For instance, Bendersky et al. \cite{BenderskyCD11}
estimate the quality of web documents by features indicating readability, layout and ease-of-navigation, such as: 
the number of terms that are rendered visible by a web browser; 
the number of terms in the title; 
the average number of characters of visible terms on the page (used also as an estimate of  
readability by \cite{KanungoO09}); 
the fraction of anchor text on the page (used also as discriminative of content by \cite{NtoulasNMF06});
the fraction of text that is rendered visible by a web browser, compared to the full source of the page (also known  
as information-to-noise ratio and used as feature of document quality in \cite{ZhouC05});
the stopword/non-stopword ratio of the page; and
the lexical entropy of the page, computed over the terms occurring in the document. 
In fact, Bendersky et al. use this type of lexical entropy as an estimate of document cohesiveness, reasoning that
 documents with lower entropy will tend to be more cohesive
and more focussed on a single topic. This inversely proportional relation between text entropy and its cohesiveness is also 
used in our model; the difference is that we compute the discourse entropy (entropy of discourse entity $n$-grams), whereas Bendersky et al.
compute the lexical entropy (entropy of individual words). 

%

Tan et al. \cite{TanGP12} also present a model of text comprehensibility using lexical features, such as bag of words, word length and sentence length, 
to approximate 
semantic and syntactic complexity. They build a classifier that uses these features to assign a comprehensibility score to each document,
 and then rerank retrieved documents according to 
this comprehensibility score. We also rerank results using our coherence scores, and like \cite{TanGP12}, find this to be effective. 
The difference of our work to that of Tan et al. is that (i) our coherence scores are not produced by a classifier 
but they are computed either as entropy or graph centrality approximations without tuning parameters; 
and (ii) we do not use lexical frequency statistics, but solely discourse entities, e.g. subject, object, which we consider better approximations of 
semantic complexity than lexical frequency features.


%
\section{Model I: Discourse Entropy for Document Coherence}~\label{sec:ecohmodel}
Our first coherence model uses information entropy. 
Entropy has been used in various areas e.g., lossless data compression~\cite{huffman1952method}, or cryptography~\cite{cachin1997entropy}. 
See Berger et
al.~\shortcite{berger1996maximum} for an older but comprehensive introduction to entropy for NLP. 

Information entropy is the expected value of the information content of a random variable. 
If a document is seen as a sequence of $N$ i.i.d.\ events $a_1 a_2 \cdots a_N$, the entropy is, roughly,
a measure of the average ``surprise'' of observing an event $a_i$. For example,
if all events occur with equal probability, the entropy is \emph{high}; however if a single event
occurs much more frequently than others, the entropy is \emph{low} (the average surprise is low as a single event will occur very often, as expected).

For a positive integer $n$, we may consider the probability $p(a_i \vert a_{i-1} \cdots a_{i-n})$
of observing event $a_i$ given the preceding $n$ events. In this case, the entropy, roughly, measures
the average surprise of seeing event $a_i$ given the history of $n$-preceding events (a so-called $n$-\emph{gram}). In particular,
if the events are \emph{discourse entities}, 
\emph{low} entropy will occur if only a few distinct discourse entities can occur, on average, after each
$n$-gram of other discourse entities.  
In contrast, if new discourse entities are being introduced
throughout the text independently of the preceding entities, e.g., recall the example in Figure \ref{fig:cat}, high entropy will occur. Based
on these observations, we compute a document coherence score as the reciprocal entropy of a random
variable constructed from the probability of discourse entities in the document. We next describe how we compute this probability of discourse entities 
(Section \ref{ss:prob}) and the final coherence score per document (Section \ref{ss:score}).

\begin{table}
\small
\centering
\caption{\footnotesize Entity grid example. Discourse entities: subject (s), object (o). 1-5 are sentence numbers.}
\label{tbl:examplegrid}
\scalebox{0.92}{
\begin{tabular}{l}
\hline
$\;\;\;\;\;\;\;\;\;\;\;\;\;\;\;\;\;\;\;\;\;\;\;\;\;\;\;\;\;\;\;\;\;\;\;\;\;\;\;\;\;$\textbf{SAMPLE TEXT}\\
\hline
1 ``One", the old man said; his hope and his confidence had never gone. \\ 
2 ``Two", the boy said. \\ 
3 ``Two", the old man agreed; ``you didn't steal them?" \\ 
4 ``I would", the boy said, ``but I bought these.". \\
5 ``Thank you", the old man said. \\
$\;\;\;\;\;\;\;\;\;\;\;\;\;\;\;\;\;\;\;\;\;\;\;\;\;\;\;\;\;\;\;\;\;\;\;\;\;\;\;\;\;$\\
\hline
$\;\;\;\;\;\;\;\;\;\;\;\;\;\;\;\;\;\;\;\;\;\;\;\;\;\;\;\;\;\;\;\;\;\;\;\;\;\;\;\;\;$\textbf{ENTITY GRID}\\
\hline
\end{tabular}
}
\scalebox{1.0}{
\begin{tabular}{c|p{0.01\columnwidth}p{0.01\columnwidth}p{0.01\columnwidth}p{0.01\columnwidth}p{0.01\columnwidth}p{0.01\columnwidth}p{0.01\columnwidth
}p{0.01\columnwidth}}
&\rotatebox{90}{\scriptsize MAN}&\rotatebox{90}{\scriptsize HOPE}&\rotatebox{90}{\scriptsize CONFIDENCE}&\rotatebox{90}{\scriptsize
BOY}&\rotatebox{90}{\scriptsize YOU}&\rotatebox{90}{\scriptsize THEM}&\rotatebox{90}{\scriptsize I}&\rotatebox{90}{\scriptsize THESE}\\[0.5em]\hline
&&&&&&&&\\[0.5em]
1 &\sub &\sub &\sub &\miss&\miss&\miss&\miss&\miss\\
2 &\miss&\miss&\miss&\sub &\miss&\miss&\miss& \miss\\
3 &\sub &\miss&\miss&\miss&\sub &\obj &\miss& \miss\\
4 &\miss&\miss&\miss&\sub &\miss&\miss&\sub & \obj\\
5 &\sub &\miss&\miss&\miss&\obj&\miss&\miss& \miss\\
\end{tabular}
}
\end{table}

%
%
\subsection{Discourse entities and their probabilities}
\label{ss:prob}
We build an entity grid as per Barzilay and Lapata \shortcite{Barzilay:2008}: 
the rows correspond to the sentences of a document $d$, in order, and each column corresponds to a salient entity occurring in the sentence, 
in order of their occurrence in $d$.
The entry in each row and column of the grid 
is the syntactic role of the corresponding salient entity. 
We use the syntactic roles of \texttt{subject} (\sub) and \texttt{object} (\obj) because they are the most important discourse items \shortcite{Barzilay:2008}, denoting respectively the \textit{actor} and the \textit{entity that is acted upon}. 
Table~\ref{tbl:examplegrid} displays an example of the entity grid built from an excerpt of Hemmingway's \texttt{"The Old Man and the Sea"}. For instance, 
\texttt{MAN} is a discourse entity occurring as a subject in sentences $1$, $3$, and $5$.

%
%
%
%
%

%
%
We propose to measure coherence using entropy in the following way. From the entity grid, we extract $n$-grams of entities in the order 
they occur in the
text, i.e. per row in the grid. We then compute the probability of each entity $n$-gram using the standard maximum likelihood language modelling 
frequency approximations (more elaborate approximations are also possible, see for instance \cite{hinrich}, but we choose this for simplicity in this preliminary work). 
That is, for an 1-gram, we compute the probability $p(e_i)$ of a discourse entity $e_i$ in document $d$ as:
\begin{equation}
\label{eq:1gram}
p(e_i) = \frac{f(e_i)}{\vert E\vert}
\end{equation}
\noindent where $f(e_i)$ is the frequency of $e_i$ in $d$,
and $\vert E \vert$ is the total frequency of all discourse entities in $d$. Similarly, for a 2-gram,
we compute the probability $p(e_i \vert e_{i-1})$ of entity $e_i$ following entity $e_{i-1}$ 
as:
\begin{equation}
\label{eq:2gram}
p(e_i \vert e_{i-1}) = \frac{f(e_{i-1},e_i)}{f(e_i)} 
\end{equation}
\noindent where $f(e_{i-1},e_i)$ is the number of times that entity $e_i$ occurs as the first entity after entity $e_{i-1}$ in $d$. 
The resulting
probability distributions may be smoothed by standard methods (e.g., Dirichlet or Good-Turing). For simplicity we do not do so in this basic model.
We use these probabilities to compute entropy as explained next.

\subsection{Entropy and coherence scores}
\label{ss:score}
Formally, the entropy, $H(X)$, of a discrete random variable variable $X$
with sample space $\Omega$ is: 
\begin{equation}\label{eqn:orgentropy}
H(X) = -\sum_{x\in\Omega} p(x)\log_2{p(x)}
\end{equation}
where $p(x) = \mathrm{Pr}(X=x)$ for all $x \in \Omega$. 
We model each document as a Markov process generating discourse entities. For a $k$-order Markov process,
the probability of generating a discourse entity $e_i$ depends solely on the preceding $n$-gram of discourse entities.
Using Equation~\ref{eqn:orgentropy} it is straightforward to obtain explicit expressions for the entropy
of the probability distribution generating events. For instance, for $k=0,1$, the resulting expressions are, respectively:
\begin{align}
H_0(E) &= -\sum_{e_i\in\Omega} p(e_i)\log_2{p(e_i)} \label{eqn:e0def}\\
H_1(E) &= -\sum_{e_i\in\Omega} p(e_i)  \sum_{e_{i-1} \in \Omega} p(e \vert e_{i-1}) \log_2 p(e \vert e_{i-1}) \label{eqn:e1def}
\end{align}
where $\Omega$ is the set of all discourse entities $E$ in the document, $p(e_i)$ is computed using Equation \ref{eq:1gram}, and $p(e_i \vert e_{i-1})$ is 
computed using Equation \ref{eq:2gram}. Subscripts $_0,_1$ denote the order of the entropy model, which increases as the value $n$ of the $n$-gram increases.  

Our document coherence score is then the reciprocal of this entropy value:
\begin{equation}
C = \frac{1}{H_k(E)}
\end{equation} 
\noindent to boost low entropy scoring documents according to our assumption that coherence and entropy are inversely related. 
The value of $C$ is always non-negative, but has no upper bound. So we normalise it as follows:
\begin{equation}
\label{eq:coh}
COH(H_k(E)) = \frac{C}{C_{max}}
\end{equation}
where $C_{max}$ is the maximum value of $C$ across all documents in the collection for a $k$-order model. 
The division by $C_{max}$ is a simple normalisation of 
the coherence scores of the documents in the collection. 
Other normalisation approaches can also be used, which, if parameterised, can result in better performing models than the ones we report. In this preliminary work, 
we use this basic normalisation.   
In addition, different values of $k$ give different orders of the Markov process, hence different coherence model instantiations, and hence distinct coherence scores. 
In this work, we experiment with 0-order, 1-order and 2-order models, as explained in Section \ref{s:eval}.

Our entropy coherence model captures only local coherence
through entity transitions occurring within sentences. We next present a family of models that capture
both local and global coherence
through entity transitions between both adjacent and non-adjacent sentences in a document.

\section{Model II: Discourse Graph Metrics for Document Coherence}\label{sec:graphcomodel}
\begin{figure}[!ht]
\centering
\scalebox{0.70}{
\includegraphics[scale=1.0]{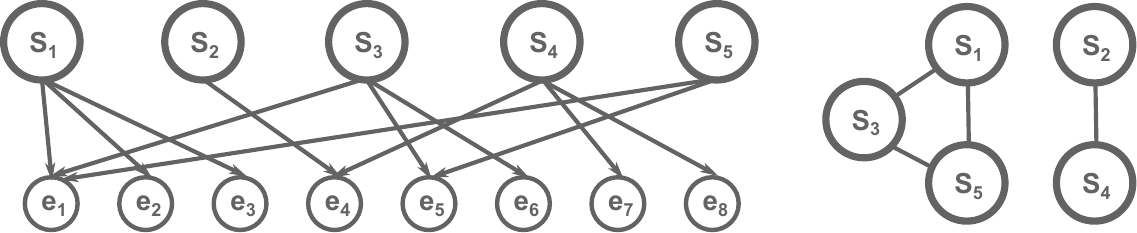}
}
\caption{Bipartite graph of the entity grid in Table~\ref{tbl:examplegrid} (left) and the corresponding (unweighted) undirected graph (right). $S$
marks sentence nodes and $e$ entities extracted from the entity grid.}
\label{fig:bipartitegraph}
\end{figure}

We represent a document $d$ as a directed bipartite graph where each node in the first partition is a sentence, 
and each node in the second partition is a discourse entity. There is an edge from a sentence-node to an entity-node
if{f} the sentence contains the entity. This representation was first suggested by Guinaudeau \& Strube~\shortcite{Guinaudeau:Strube:2013}. Using this
directed bipartite graph, we can build an undirected
graph whose nodes are sentences and where there is an edge between two distinct nodes if{f} they share at least one common entity. This undirected graph can be unweighted or weighted; if weighted, 
the weight can reflect salient properties of the document and sentences, e.g. the distance of the sentences in the text. An example 
of a bipartite graph and its corresponding undirected graph is shown in
Figure~\ref{fig:bipartitegraph}.

The intuition is that as these graphs model relationships between sentences and entities, properties of the graphs will reflect the coherence of a
document, as in coherent text, sentences occurring close to each other should have closely related entities.
Guinaudeau and Strube \shortcite{Guinaudeau:Strube:2013} use only the outdegree of such graphs constructed from discourse entities to calculate a coherence
score.  Note that
while Guinaudeau and Strube write that their model only takes \emph{local} coherence into account,
the bipartite graph can also be used to reason about \emph{global} coherence, as it models connections between non-adjacent sentences.
We extend the approach of Guinaudeau and Strube by experimenting with a number of other graph metrics that capture
various aspects of the topology of the graphs (and accordingly, we conjecture, the narrative flow of the text). Our methodology consists of two main
steps: (i) we build, for each $d$, a discourse entity graph (Section \ref{ss:build}), and (ii) we compute our proposed graph topology measures 
and use these as coherence
scores for each document $d$ (Section \ref{ss:metrics}). We next describe these steps in detail.

%
%
%
%
%
%
\subsection{Building discourse entity graphs}
\label{ss:build}
For a document $d$ containing $N$ sentences $s = \{s_i,\ldots,s_N\} : i = 1,\ldots,N-1$, we denote by $\hat{G}_d = (V,U,E)$ the labelled directed
bipartite graph with $V = \vert N\vert$ and $U$ nodes where $v\in V$, $u\in U$, $V\cap U = \emptyset$. $E = \{(v,u) : V\times U\}$ consists of
ordered pairs of nodes from $V\times U$. A node $v\in V$ denotes a sentence and $u\in U$ an entity found in the entity grid that can be shared by
multiple nodes in $V$. $\hat{G}_d$ is called the \emph{bipartite graph} of $d$. The labelling of the nodes is used to retain the \emph{order} in which
the sentences
represented by nodes occur in $d$.

We denote by $G_d = (V,E)$ the undirected graph with $V$ nodes and $E$ edges where $v,u\in V$ are nodes in $V$ and $(v,u)\in E$ is an edge between
pairs of nodes $v,u\in V$ with a non-zero real-valued weight. A node $v\in V$ denotes a sentence in $d$ and an edge $(v,u)\in E$ between nodes $v,u\in
V$ corresponds to an entity shared by $v$ and $u$. 
$G_d$ is called the \emph{projection graph} of $d$. 

%
%
\subsection{Using graph metrics for coherence scores}
\label{ss:metrics}
We use several graph metrics applied to the graphs $\hat{G}_d$ and $G_d$ associated to document $d$. Each metric produces a separate coherence score for $d$. 
We present these next.
%
\subsubsection{PageRank}\label{metric:pagerank}
\textit{PageRank}~\cite{brin:1998} is a vertex ranking metric that gives higher scores to the best connected vertices in a graph. The PageRank score
of a node in a graph depends on its indegree and the PageRank scores of those nodes linking to it (the latter being considered as recommendation).
Formally, the PageRank of $G_d$ is given by:

\begin{equation}
PR = c\displaystyle\sum_{v\rightarrow u}\frac{1}{d_v}PR(u) + (1 - c) 
\end{equation}

\noindent where $\mathrm{PR}(u)$ is the PageRank of $u$, $d_v$ is the number of edges incident on node $v$, and $c$ is a damping factor (typically $c
= 0.85$). We hypothesise that a lower median PageRank score is indicative of a more coherent document: as $G_d$ becomes more connected, the relative
importance of each node or sentence decreases in the PageRank score. We use the median rather than the average because the average PageRank score does not
distinguish between star-graphs (where all but one node are connected to the single remaining node, and no other edges occur) and path graphs (where
the nodes occur in sequence); however, path graphs intuitively correspond to coherent discourse flow whereas star graphs do not.

%
So we propose that the final coherence score of $d$ is the \emph{median} PageRank of nodes of $G_d$:

\begin{equation}\label{eqn:pagerank2}
COH_{PR} =
\mathrm{median}_{u \in V}\left( c\displaystyle\sum_{v\rightarrow u}\frac{1}{d_v}PR(u) + (1 - c) \right) 
\end{equation}

\subsubsection{Clustering Coefficient}
The clustering coefficient (CC) measures the extent to which the neighbours of a node in $G_d$ are connected to each other. We hypothesise that a
lower clustering coefficient score is indicative of a more coherent document: the fact that the neighbours of any given node in $G_d$ are themselves
connected suggests that this node's importance to the overall discourse is very low; if no neighbours are connected, all nodes are equally
important for coherence. 

We compute the coherence score of $d$ as the global clustering coefficient of $G_d$:
\begin{equation}\label{eqn:ccdef}
COH_{CC} = \frac{1}{\vert V\vert} \displaystyle\sum_{u\in V} \frac{\delta(u)}{\tau(u)}
\end{equation}
where $\delta(u)$ is the number of closed triplets containing $u$, and $\tau(u)$ is the total number of open and closed triplets containing $u$ (a
\emph{triplet}
is a subset of $V$ containing exactly three nodes. A triplet is \emph{open} if exactly two of its three nodes are connected,
and \emph{closed} if all three nodes are connected).

\subsubsection{Betweenness}
The betweenness of a node $u$ measures the fraction of shortest paths in $G_d$ that contain $u$.
We hypothesise that a higher average betweenness score for the nodes of $G_d$ is indicative of a more coherent document: a coherent document $d$ in our
context should resemble a path graph where a sentence is connected only to the preceding and next sentence. 

We compute the coherence score of $d$ as the average of all betweenness scores in $G_d$:
\begin{equation}\label{eqn:betweenness}
COH_{BW} =  \frac{1}{\vert V\vert}\displaystyle\sum_{\forall u\in V} \displaystyle\sum_{s\in V}\sum_{t\in V\setminus
s}\frac{\beta_{s,t}(u)}{\beta_{s,t}}
\end{equation}
where $\beta_{s,t}(u)$ is the number of shortest paths between $s$ and $t$ in $V$ containing $u$, and $\beta_{s,t}$ is the total number of shortest
paths between $s$ and $t$.
%
%
%
\subsubsection{Entity distance}
The entity distance between two sentences $u$ and $v$ is the smallest number of words occurring between
$u$ and $v$ that do not contain an entity shared by $u$ and $v$. We reason that a short entity distance implies
that shared entities between $u$ and $v$ aid the flow of the text, whereas a high distance means that
many other entities are mentioned in between $u$ and $v$, implying topic drift, and hence lower coherence.

We compute the coherence score of $d$ as the inverse of the average entity distance over all entities shared by two or more sentences:
\begin{equation}\label{eqn:entitydistance}
COH_{ED} = \left(\frac{1}{\vert S\vert}\displaystyle\sum_{\forall e \in \mathcal{E}} \displaystyle\sum_{i=1}\sum_{j=i+1} l_{j}(e) -
l_{i}(e)\right)^{-1}
\end{equation}
where $\vert S\vert$ is the number of sentences in $d$, $\mathcal{E}$ is the set of entities of $d$ occurring in at least two distinct sentences,
and $l_{i}(e)$ and $l_{j}(e)$ are the locations of entity $e$ in the document. For example, for sentence $s_i$, the location of entity $e$ could be the 20th term in
$d$, and for $s_{i+1}$, $e$ could be the 30th term in $d$.

\subsubsection{Adjacent Topic Flow (ATF)}\label{subpar:atf}
We propose an adjacent topic flow (ATF) metric, which is, roughly, the average of the reciprocal
number of shared entities between adjacent sentences. In a coherent document, consecutive sentences should share common entities between them to
relate current discourse to past discourse: the fewer shared entities (minimum one) between adjacent sentences, the more focused the discourse and the
more coherent the text. We compute the coherence score of $d$ as the average reciprocal of the union of entities between adjacent sentences:
\begin{equation}\label{eqn:atf}
COH_{ATF} = \frac{1}{N-1}\displaystyle\sum_{(s_i,s_{i+1})} \frac{1}{\sigma(s_i,s_{i+1})}
\end{equation}
where $s_1,\ldots,s_N$ are the sentences in the document and $\sigma(s_i,s_{i+1})$ is the union of entities from sentences $s_i$ and $s_{i+1}$. The
intuition is that adjacent sentences should share entities, but if the union of their entity sets is large this strains the focus of the reader as
these entities need not be necessarily linked to previous discourse.

\subsubsection{Adjacent Weighted Topic Flow (AWTF)}
The ATF metric determines if a pair of adjacent sentences share a single entity irrespective of whether several entities are shared. Thus, a text with
two (or more) main topics (e.g.\ \texttt{security} and \texttt{heartbleed}) mentioned in adjacent sentences would not be detected. For coherence, more
shared entities can be seen as more salient words aiding the reader to retain focus as more links exist between current and past discourse. We present
a version of ATF called Adjacent Weighted Topic Flow (AWTF) where we \emph{weigh} a document according to the number of entities shared by adjacent
sentences. The resulting coherence score is:
\begin{equation}\label{eqn:awtf}
COH_{AWTF} = \frac{1}{N-1}\displaystyle\sum_{(s_i, s_{i+1})}\omega(s_i,s_{i+1})
\end{equation}
\noindent where $\omega(s_i,s_{i+1})$ is the number of shared entities. 

\subsubsection{Non adjacent Topic Flow Metrics}
\begin{figure}[!ht]
\centering
\scalebox{1.0}{
\includegraphics[trim=3mm 0mm 0mm 0mm ,clip,scale=0.6]{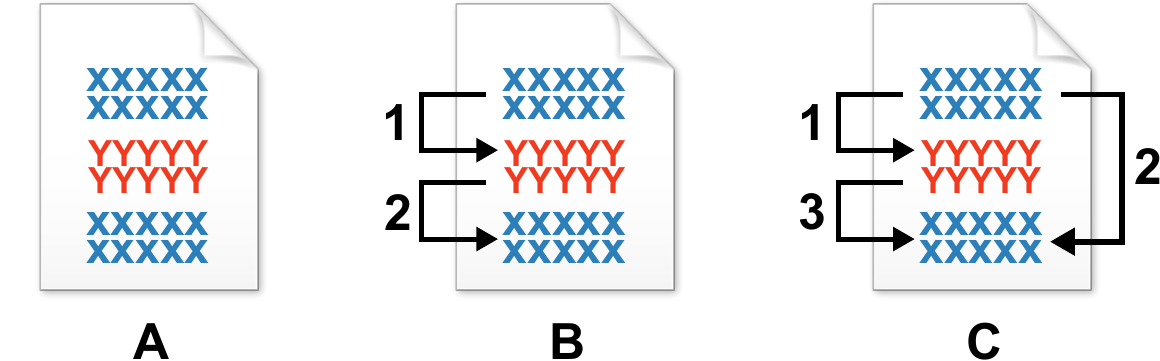}
}
\caption{(A) Document with three paragraphs where two are on the same topic. X and Y indicate topics. (B) Application of an adjacent topic flow metric. (C) Application of an
non adjacent topic flow metric. Numbers indicate order of comparison.}
\label{fig:natfmetrics}
\end{figure}

ATF and AWTF rely  on \emph{local} coherence between adjacent sentences to capture global coherence. However,
a coherent text may contain discourse gaps where several topics are treated locally, then abandoned, yet later on picked up again. For example,
consider a document with three paragraphs of which two are on the same topic as illustrated in Figure~\ref{fig:natfmetrics}.
Figure~\ref{fig:natfmetrics}(B) shows that the application of an adjacent-based topic flow metric would find no coherence between the first two and
last two sentences. Conversely, Figure 1(C) shows that a \emph{non}-adjacent topic flow metric would find some coherence between the first and last
sentence in the second comparison (indicated by the number 2) as they are on the same topic.

We define two coherence models (i) \emph{non-adjacent topic flow (nATF)} and (ii)
on \emph{non-adjacent weighted topic flow (nAWTF)}, with corresponding coherence scores:
\begin{equation}\label{eqn:natf}
COH_{nATF} = \frac{1}{\hat{E}}\displaystyle\sum_{s_i}\sum_{s_j} \frac{1}{\sigma(s_i,s_{j})}
\end{equation}
and
\begin{equation}\label{eqn:nawtf}
COH_{nAWTF} = \frac{1}{\hat{E}}\displaystyle\sum_{s_i}\sum_{s_{j}} 
\left\{
     \begin{array}{ll}
       1 & :\hat{e}\in s_i\land s_{j}\\
       0 & :\mbox{otherwise}
     \end{array}
   \right.
\end{equation}
where $\hat{E}$ is the number of all entities in $d$ shared between at least two sentences. If $\hat{E} = 0$, we set $COH_{nAWTF} = 0$ reflecting that
there is no coherence in the document. 

Strictly speaking, Equations \ref{eqn:atf}--\ref{eqn:nawtf} are not graph centrality metrics but as our bipartite graph is a labelled graph all are
invariant under graph isomorphism.

We next evaluate the above graph-based models, as well as the entropy models of document coherence presented in Section \ref{sec:ecohmodel}.
\section{Evaluation}
\label{s:eval}

First we evaluate the accuracy of our coherence models in Section \ref{ss:accuracy}, and then retrieval precision when reranking documents according to
their coherence scores in Section 
\ref{ss:precision}. 

\subsection{Experiment 1: Coherence Model Accuracy}
\label{ss:accuracy}
\subsubsection{Experimental Setup}
We evaluate our coherence models in the \emph{sentence reordering} task, which is standard for coherence evaluation. The main idea is that we scramble
the order of the 
sentences in each document, and subsequently then 
determine the number of times the original document is deemed more coherent than its permutations.
Specifically, for each original document $d_o$ in a dataset, we reorder its sentences 20 times, producing 20 permutations $d_i : i \in [1,
\ldots, 20]$. 
We assume that the more sentences are reordered, the less coherent $d_i$ becomes on average; this assumption has been validated with human
assessments \cite{LinNK11}. 
We run our coherence model on all $(d_o,d_i)$ pairs, and measure accuracy by considering as \textit{true positive} each case where $d_o$ gets a higher
coherence score 
than its $d_i$ permutation.

We use the earthquakes and accidents dataset\footnote{\url{http://people.csail.mit.edu/regina/coherence/}}, which has been used in previous work on coherence prediction~\cite{Barzilay:2008}, so that we can
directly compare our coherence model
 to the state of the art. This dataset contains relatively clean, curated articles about earthquakes from the North American News Corpus and 
 narratives from the National Transportation Safety Board, of relatively short length and not much variation in document length (statistics in Table
\ref{tbl:datasets}). 
We identify entities in documents using the Stanford parser\footnote{\url{http://nlp.stanford.edu/software/lex-parser.shtml}}. Due to the steep
increase in parsing time as sentence 
length increases, we only consider sentences of 
60 terms or less, which is approximately 3 times the average length of a sentence in English~\cite{sigurd2004word}. If a sentence is
longer than 60 terms, we do not 
exclude it, but rather cut it at length 60, to speed up processing. 

Our baselines are (i) Barzilay and Lapata's original entity grid model \cite{Barzilay:2008}, and (ii) Barzilay and Lee's well-known HMM-based model
\cite{BarzilayL04}. 
We do not use Guidnaudeau and Strube's indegree model \cite{Guinaudeau:Strube:2013} as they use a different dataset. 
For the remaining baselines, we compare the \emph{tuned} scores reported in \cite{Barzilay:2008} with \emph{untuned} scores of our models.


\begin{table}
\centering
\caption{Statistics of the earthquakes and accidents dataset. Average document length is measured in terms; MAD denotes mean absolute deviation.}
\label{tbl:datasets}
\begin{tabular}{l|c|c}
\hline
& Earthquakes & Accidents\\
\hline
\# documents & 100 & 100 \\
average doc. length& 257.3 & 223.5 \\
MAD doc. length & 101.8 & 31.9  \\
\hline
\end{tabular}
\end{table}

\subsubsection{Sentence Reordering Findings}
Table~\ref{tbl:acc} presents the accuracy and statistical significance of the sentence reordering experiments. 
Our models are overall comparable to the baselines. Our 0-order entropy model outperforms both baselines for both datasets. As 
its order increases, its performance decreases 
(but still outperforms both baselines for the accidents dataset). This happens because as the order of our model grows, we expect to find fewer
higher-frequency $n$-grams
 because our model will rely on increasingly longer sequences of entities. 
This results in overall higher entropy scores. For example, a 2-gram and 3-gram coherence model of the example entity grid in 
Table~\ref{tbl:examplegrid} would result in these respective
entropy scores $H(2) = 3.2776$ 
and $H(3) = 3.3219$, which grow as the value of $n$ in the $n$-gram increases.

Looking at cases where our entropy model errs, we see that there is a potentially negative bias of our model against shorter documents. 
As our entropy model scores document coherence using the frequency of entity transitions that occur in a document, short documents are at a
disadvantage. For example, a very short document of only one sentence where all $n$-grams occur exactly once will receive a very high entropy score (hence very low coherence score). 
The minimum number of sentences found in documents in our data were 2 and 3 for the earthquakes and accidents datasets respectively. 
Our entropy model will very likely deem these documents as incoherent, even if they are not.

Regarding our graph coherence metrics, the PageRank, betweenness, entity distance, ATF, nATF and nAWTF variants also outperform the baselines at all times. 
Entity distance and betweenness are overall best, outperforming our entropy model too. 
Both entity distance and betweenness measure the same feature, that is distance between discourse entities, but in different ways: betweenness 
in terms of shortest path in the graph; entity distance in terms of absolute number of terms separating the entities. It seems that this distance, 
measured in either way, is more discriminative of coherence than e.g. any clustering or recommendation of discourse entities captured by the 
clustering coefficient or PageRank respectively.

Overall, manual inspection of the documents where most of our models failed reveals two main reasons for this: 
(I) These documents contain sentences of very large length, exceeding the 60 term limit we have defined. 
As we artificially cut these sentences at 60 terms, the discourse flow of the text is disturbed. 
Moreover, several very long sentences tend to introduce a large number of new entities into the discourse, 
which may be topically relevant but not necessarily identical, e.g. 
$\dots$\textit{accommodation with a bed, table, cupboard, attached washroom, television, broadband Internet facility, telephone, small kitchen 
with facilities to prepare tea, coffee, etc.} 
(II) The versions of our models that concentrate on entities shared by adjacent sentences risk underperforming, 
because about 50\% of the sentences in English are estimated not to share any entities and 60\% of them to be related 
by weak discourse relations~\cite{louis:2010}.  E.g., for journalistic texts like the earthquake dataset, the spatial 
proximity of sentences does not necessarily correlate with their semantic relatedness, as related sentences can be placed 
at the beginning and the end for emphasis \cite{zhang:2011}. 
Indeed in Table \ref{tbl:acc} our two adjacent topic flow metrics ATF and AWTF have lower accuracy in the earthquakes than accident dataset, and also compared to their non-adjacent versions (nATF and nAWTF).
Note that this spatial proximity limitation is a major disadvantage of \textit{all} models of local text coherence, not only ours. We discuss this point in Section \ref{s:discussion}.

Overall, our \emph{untuned} models perform on par with the \emph{tuned} baselines, occasionally outperforming them.
Motivated by the good performance of our coherence models, we next study their potential usefulness to retrieval.

 %
 %
 %
 %
 %
 %
\begin{table}
\centering
\caption{Coherence accuracy in the sentence reordering task. 
$\pm$\% is the difference from the strongest baseline and bold means $\ge$ strongest baseline. 
* marks statistical significance at the $0.05$ interval using the paired $t$-test and $\ddagger$ means best overall.}  
\label{tbl:acc}
\scalebox{0.75}{
\begin{tabular}{c|l|lr|lr}
\hline
&Method &\multicolumn{2}{c}{Earthquakes} &\multicolumn{2}{c}{Accidents} \\
																	&&Acc.&$\pm$\%&Acc.&$\pm$\%\\
\hline
\multirow{2}{*}{BASELINES}&Entity grid model \cite{Barzilay:2008}	&69.7\%*&--		&67.0*&--\\
&HMM-based model \cite{BarzilayL04}									&60.3*&--		&31.7*&--\\
\hline
\multirow{3}{*}{ENTROPY}&Entropy-0 order							&\bf75.0 &+7.6\%		&\bf73.0*	&+9.0\%\\
&Entropy-1 order 													&64.0	&--8.2\%		&\bf70.0*	&+4.5\%\\
&Entropy-2 order 													&64.0	&--8.2\%		&\bf70.0*	&+4.5\%\\
\hline
\multirow{8}{*}{GRAPH}&PageRank										&\bf75.0 &+7.6\%		&\bf73.0*	&+9.0\%\\
&Clustering Coef.												&67.0	&--3.9\%		&66.0* &--1.5\%	\\
&Betweenness														&\bf73.0*	&+4.7\%	&$\ddagger$\bf77.0* &+14.9\%	\\
&Entity Distance													&$\ddagger$\bf76.0 &+9.0\%		&\bf75.0*	&+11.9\%\\
&Adj. Topic Flow													&\bf70.0*	&+0.4\%	&\bf74.0*	&+10.4\%\\
&Adj. W. Topic Flow													&61.0*&--12.5\%		&66.0* &--1.5\%	\\
&nAdj. Topic Flow													&\bf70.0 &+0.4\%	&\bf70.0 &+4.5\%	\\
&nAdj. W. Topic Flow												&\bf70.0 &+0.4\%		&\bf70.0*	&+4.5\%\\
\hline
\end{tabular}
}
\end{table}

\subsection{Experiment 2: Retrieval with Coherence}
\label{ss:precision}

\subsubsection{Experimental Setup}
We now test whether our document coherence scores can be useful to retrieval. 
Our assumption is that more coherent documents are likely to be more relevant. 
To test this we rerank the top 
1000 documents retrieved by a baseline model according to their coherence scores. Our baseline ranking model is a unigram, query likelihood,
Dirichlet-smoothed, language model. 
Let $RSV$ be the baseline retrieval status value of a document, and  
$COH$ be the coherence score of a document computed as per any of our coherence models (Equations \ref{eq:coh}--\ref{eqn:nawtf}). 
For our entropy coherence model we use only the 0-order variant, as this performed best in the sentence reranking task. 
We use a simple \textit{linear} combination \cite{Cras05} to compute the reranked $RSV$ of each document, denoted $\widehat{RSV}$:
\begin{equation}
\label{eq:linear}
\widehat{RSV} = RSV \times \alpha + COH \times (1-\alpha)
\end{equation}
\noindent where $0 \le \alpha \le 1$ is a smoothing parameter controlling the effect of $RSV$ over $COH$. 

We use Indri 5.8\footnote{http://www.lemurproject.org} for indexing and retrieval without stemming or stopword removal. 
We use the ClueWeb09 cat.B\footnote{http://www.lemurproject.org/clueweb09.php/} test collection with queries 150-200 from the Web AdHoc track of TREC
2012. 
ClueWeb09 contains free text crawled from the web, likely to be more noisy and containing documents that are overall longer and
with more variation in document 
 length than in the earthquakes and accidents dataset. We remove spam from ClueWeb09 using the spam rankings of Cormack et al.
\shortcite{fusion} 
 with a percentile-score $<90$ indicating spam. This threshold is stricter than the one recommended in \shortcite{fusion}, practically meaning that 
 we remove many more documents assumed to be spam. 
 This reduces the number of documents from ca. 50 million to ca. 16 million. We apply such strict spam filtering for the following reason:
as we hypothesise that a lower entropy scoring document will be more coherent, documents containing the same
repeated entities will be judged more coherent. For example, a document containing only the sentence \texttt{free domains, free domains, free domains}
would receive an entropy score of $0$ (i.e., highest coherence) but is, arguably, not very coherent. While this problem does not occur in the earthquakes and accidents
 dataset, spam documents are
plentiful in ClueWeb09~\cite{fusion} which is why apply such a strict threshold when removing spam from this collection.  
We evaluate retrieval with Mean Reciprocal Rank (MRR) of the first relevant result, Precision at 10 (P@10), Mean Average Precision (MAP) of the top 1000 results, and 
Expected Reciprocal Rank at 20 (ERR@20). 

The baseline and our reranking method include parameters $\mu$ and $\alpha$ that we tune using 5-fold cross-validation. We report the average of the five test folds. 
We vary the ranking baseline's $\mu \in$ \{100,500,800,1000,2000,3000,4000,5000,8000,\\10000\}; 
and the reranking parameter $\alpha \in \{0.5..1\}$ in steps of 0.05.

\subsubsection{Retrieval Findings}


\begin{table}
\centering
\caption{\label{tab:retrieval} Retrieval precision without and with coherence.
$\pm$\% is the difference from the strongest baseline and bold means $\ge$ strongest baseline. 
}
\scalebox{0.85}{
\begin{tabular}{l|l|lr|lr}
\hline
\multicolumn{2}{c|}{Method}				&MRR&$\pm$\%		&P@10&$\pm$\%	\\
\hline
\multicolumn{2}{c|}{Baseline} 			&20.57&				&19.80&		\\
\hline
\multirow{9}{*}{\rotatebox[origin=c]{90}{COHERENCE}}
&Entropy-0 order		&\bf49.50&+140.6\%	&\bf33.00&+66.7\%	\\
&PageRank			&\bf49.85&+142.3\%	&\bf34.40&+73.7\%	\\
&Clustering Coef.	&\bf51.82&+151.9\%	&\bf34.60&+74.7\%	\\
&Betweeness			&\bf49.74&+141.8\%	&\bf36.40&+83.8\%	\\
&Entity Distance		&\bf34.18&+66.2\%	&\bf22.40&+13.1\%	\\
&Adj. Topic Flow		&\bf55.73&+170.9\%	&\bf34.20&+72.7\% \\
&Adj. W. Topic Flow	&\bf51.60&+150.8\%	&\bf34.20&+72.7\%	\\
&nAdj. Topic Flow	&\bf50.62&+146.1\%	&\bf34.40&+73.7\%	\\
&nAdj. W. Topic Flow&\bf50.79&+146.9\%	&\bf34.60&+74.7\%	\\
\hline
\hline
\multicolumn{2}{c|}{Method}				&MAP&$\pm$\%		&ERR@20&$\pm$\%\\
\hline
\multicolumn{2}{c|}{Baseline}  			&10.07&				&11.78&\\
\hline
\multirow{9}{*}{\rotatebox[origin=c]{90}{COHERENCE}}
&Entropy-0 order		&09.79&	--2.8\%		&\bf20.18&+71.3\%\\
&PageRank			&\bf10.12&+0.5\%	&\bf21.15&+79.5\%\\
&Clustering Coef.	&\bf10.20&+1.3\%	&\bf21.19&+79.9\%\\
&Betweeness			&\bf10.08&+0.1\%	&\bf21.98&+85.6\%\\
&Entity Distance		&07.22&--28.3\%		&\bf15.86&+34.6\%\\
&Adj. Topic Flow		&\bf10.12&+0.5\%	&\bf21.87&+85.6\%\\
&Adj. W. Topic Flow	&\bf10.12&+0.5\%	&\bf22.65&+92.3\%\\
&nAdj. Topic Flow	&\bf10.13&+0.6\%	&\bf21.16&+79.6\%\\
&nAdj. W. Topic Flow&\bf10.13&+0.6\%	&\bf21.15&+79.5\%\\
\hline
\end{tabular}
}
\end{table}

Table \ref{tab:retrieval} displays the retrieval results. We see that reranking documents by their coherence score overall outperforms the baseline in terms of retrieval precision, with few exceptions. 
Specifically, the MRR gains vary between +66.2\% and +170.9\%, which practically translates to a boosting of the first relevant document by moving it approximately one position higher in the ranking on average. 
The P@10 gains vary between +13.1\% and +83.8\%, which practically translates to an increase of the portion of relevant documents found in the top 10 
from less than 2 in the baseline to over 3 documents on average, except once (for entity distance). 
This gaining trend also applies to the top 20 retrieved results, as can be seen by the improvements in ERR@20. 
However, looking at MAP for the top 1000 retrieved documents we see only marginal gains and twice drops in performance (for entropy and entity distance). 
This means that 
coherence improves mainly early precision, i.e. has the potential to refine the very top of the ranked list, but not alter it significantly at 
more depth. 

Interestingly, while entity distance was among the best performing models in the sentence reranking task (Table \ref{tbl:acc}), 
it is the weakest performing model
when integrated to retrieval. This complements previous findings in the literature showing that, when using an NLP component in IR, higher NLP accuracy does not necessarily 
result in higher retrieval performance \cite{GaoNNZWH10}. 
The reason why entity distance underperforms when integrated to retrieval compared to sentence reranking could be the different dataset characteristics: 
earthquakes and accidents include relatively short 
documents (roughly 240 terms long on average) that are fairly uniform (on average 65 terms of mean absolute deviation), see Table \ref{tbl:datasets}.
ClueWeb09 on the other hand includes documents that are longer (1461 terms long on average) and more heterogeneous in style and size. 
In fact, quite 
a few of the ClueWeb09 documents are Wikipedia articles, which are fairly long and organised in thematic subsections, meaning that 
salient entities might be introduced, then dropped for a while, and later on picked up again in different subsections. 
This is likely to affect negatively entity distance, much more than our other 
coherence metrics, as it is the only metric that measures the actual distance in words between occurrences of the same entity in 
different sentences. 
So, for those documents where this distance varies wildly, reaching both very low and very high numbers, the entity distance
 will produce notably different coherence scores for different documents,
 rendering a comparison between documents on the basis of these scores difficult.

Finally, note that the baseline scores in Table \ref{tab:retrieval} are not directly comparable to the ones reported in the literature when using no spam filter or another spam filter threshold, 
because the stricter the spam filtering, the smaller the final dataset used for retrieval and the higher the risk of removing relevant documents.
 Indeed, it has been reported that for ClueWeb09 cat.B, removing spam results in overall lower precision \cite{BenderskyCD11,LiomaLL12}. We 
also anecdotally report that several documents among the ones we removed as spam were assessed as relevant in the TREC relevance judgements.

Overall we conclude that the improvements in Table \ref{tab:retrieval} show that coherence can be a discriminative feature of relevance. 

%
%
%
%
%

\section{Discussion and Future Work}
\label{s:discussion}

We now discuss interesting caveats of our coherence models and potential solutions.

One weakness of our coherence models, and of the entity grid in fact, is in capturing the contribution of newly introduced entities to discourse: it
is long assumed that repeated mentions of the same entities contribute to coherence; however, related-yet-not-identical entities also
contribute to text coherence. To our knowledge, this is something that occurs frequently in text but that is not yet captured by automatic models of coherence prediction.
 Doing so would require a richer set of
metrics and representations, involving for instance methods long used in IR for detecting near-synonyms, such as WordNet-type ontologies, or term associations extracted 
from large scale query logs or other relevant corpora. This is an interesting research direction that we intend to follow in the future.

A further limitation of our coherence models is the selection of salient discourse entities: following \cite{Barzilay:2008}, 
we consider all subjects and objects as salient discourse entities. 
This assumption implies that (a) all subjects and objects are equally salient in text, and (b) all subjects and objects are topical 
(as opposed to having a modifying or periphrastic role in the discourse). However, neither of these 
assumptions is true. Assumption (a) could be removed by including some salience weighting component in the selection of entities. In the past, this has been attempted 
through the use of linguistic features, such as modifiers and named entity types \cite{Elsner:Charniak:2011}. It will be interesting to see if we can 
also use statistical approximations to this end, looking into the rich IR literature in term weighting. 
Assumption (b) could be removed by including a similar topical weighting component, such as in \cite{ElsnerC11}. In the case where coherence scores are used for IR,
 it makes sense to align this topical detection to the query topic, so that, for instance, coherence is computed only for documents whose topic is 
 relevant to the query topic. Such a tight integration of coherence modelling to ranking would be an interesting direction to explore in future work.

\section{Conclusions}
\label{s:conc}
Modelling text coherence is an area that has received a lot of interest, with a surge of automatic methods in the last decade. 
We presented two novel classes of models
of document coherence which extend the well-known \textit{entity grid} representation \cite{Barzilay:2008}. 
Our first model approximates coherence as the
inverse of the discourse entropy in text, based on the assumption that repeated entities in text, which have low entropy, indicate higher coherence.
Our second class of coherence models map the entity grid of a document into a graph, as proposed by Guinaudeau \& Strube~\shortcite{Guinaudeau:Strube:2013}, and
then approximate document coherence using different metrics of graph topology and their variations. 
The assumption is that the discourse flow of a document can
be reflected in the topology of its entity graph, hence any regularities of the former may be measured from the latter.

We report two experiments. In the first, we find that our coherence models are comparable to two well-known text coherence models. 
Given that our models are completely untuned, their performance may be further improved through smoothing. 
In the second experiment, we find that using our coherence scores to rerank retrieved documents improves retrieval precision, especially for the top 1-20 results. 

Overall, this work contributes two novel classes of document coherence models that approximate coherence from discourse entity $n$-grams. 
Their application to retrieval complements recent work in IR showing a positive 
relation between relevance and document cohesiveness or comprehensibility \cite{BenderskyCD11,TanGP12} (albeit computed differently than we do). 
This is a promising research direction for IR that we intend to pursue in the future.

%
%
\paragraph{Acknowledgments}
Partially funded by the second author's \textit{FREJA research excellence} fellowship (grant no. 790095).



\begin{thebibliography}{10}
\begin{small}
\bibitem{Banik09}
E.~Banik.
\newblock Extending a surface realizer to generate coherent discourse.
\newblock In {\em ACL/IJCNLP (Short Papers)}, pages 305--308. ACL, 2009.

\bibitem{BarzilayEM02}
R.~Barzilay, N.~Elhadad, and K.~McKeown.
\newblock Inferring strategies for sentence ordering in multidocument news
  summarization.
\newblock {\em J. Artif. Intell. Res. (JAIR)}, 17:35--55, 2002.

\bibitem{Barzilay:2008}
R.~Barzilay and M.~Lapata.
\newblock Modeling local coherence: An entity-based approach.
\newblock {\em ACL}, 34(1):1--34, 2008.

\bibitem{BarzilayL04}
R.~Barzilay and L.~Lee.
\newblock Catching the drift: Probabilistic content models, with applications
  to generation and summarization.
\newblock In {\em HLT-NAACL}, pages 113--120, 2004.

\bibitem{BenderskyCD11}
M.~Bendersky, W.~B. Croft, and Y.~Diao.
\newblock Quality-biased ranking of web documents.
\newblock In {\em {WSDM}}, pages 95--104, 2011.

\bibitem{berger1996maximum}
A.~L. Berger, V.~J.~D. Pietra, and S.~A.~D. Pietra.
\newblock A maximum entropy approach to natural language processing.
\newblock {\em ACL}, 1996.

\bibitem{bicknell:2009}
K.~Bicknell and R.~Levy.
\newblock A model of local coherence effects in human sentence processing as
  consequences of updates from bottom-up prior to posterior beliefs.
\newblock In {\em NAACL}, pages 665--673, 2009.

\bibitem{brin:1998}
S.~Brin and L.~Page.
\newblock The anatomy of a large-scale hypertextual web search engine.
\newblock {\em CN \& ISDN}, pages 107--117, 1998.

\bibitem{burstein:2010}
J.~Burstein, J.~Tetreault, and S.~Andreyev.
\newblock Using entity-based features to model coherence in student essays.
\newblock In {\em NAACL}, pages 681--684, 2010.

\bibitem{cachin1997entropy}
C.~Cachin.
\newblock {\em Entropy measures and unconditional security in cryptography}.
\newblock Hartung-Gorre Konstanz, 1997.

\bibitem{CelikyilmazH11}
A.~\c{C}elikyilmaz and D.~Hakkani-T{\"u}r.
\newblock Discovery of topically coherent sentences for extractive
  summarization.
\newblock In Lin et~al. \cite{DBLP:conf/acl/2011}, pages 491--499.

\bibitem{ChenSB07}
E.~Chen, B.~Snyder, and R.~Barzilay.
\newblock Incremental text structuring with online hierarchical ranking.
\newblock In {\em EMNLP-CoNLL}, pages 83--91. ACL, 2007.

\bibitem{ChenBBK09}
H.~Chen, S.~R.~K. Branavan, R.~Barzilay, and D.~R. Karger.
\newblock Global models of document structure using latent permutations.
\newblock In {\em HLT-NAACL}, pages 371--379. ACL, 2009.

\bibitem{CheungP10}
J.~C.~K. Cheung and G.~Penn.
\newblock Entity-based local coherence modelling using topological fields.
\newblock In {\em ACL}, pages 186--195, 2010.

\bibitem{fusion}
G.~V. Cormack, M.~D. Smucker, and C.~L. Clarke.
\newblock Efficient and effective spam filtering and re-ranking for large web
  datasets.
\newblock {\em IR}, 14(5):441--465, 2011.

\bibitem{Cras05}
N.~Craswell, S.~Robertson, H.~Zaragoza, and M.~Taylor.
\newblock Relevance weighting for query independent evidence.
\newblock In {\em SIGIR}, pages 416--423, 2005.

\bibitem{ElsnerC08a}
M.~Elsner and E.~Charniak.
\newblock Coreference-inspired coherence modeling.
\newblock In {\em ACL (Short Papers)}, pages 41--44. ACL, 2008.

\bibitem{ElsnerC11}
M.~Elsner and E.~Charniak.
\newblock Disentangling chat with local coherence models.
\newblock In Lin et~al. \cite{DBLP:conf/acl/2011}, pages 1179--1189.

\bibitem{Elsner:Charniak:2011}
M.~Elsner and E.~Charniak.
\newblock Extending the entity grid with entity-specific features.
\newblock In {\em AMACL}, pages 125--129, 2011.

\bibitem{enos:1996}
T.~Enos.
\newblock {\em Encyclopedia of rhetoric and composition: Communication from
  ancient times to the information age}, volume 1389.
\newblock 1996.

\bibitem{FilippovaS07}
K.~Filippova and M.~Strube.
\newblock The german vorfeld and local coherence.
\newblock {\em Journal of Logic, Language and Information}, 16(4):465--485,
  2007.

\bibitem{FoltzKL98}
P.~Foltz, W.~Kintsch, and T.~Landauer.
\newblock The measurement of textual coherence with latent semantic analysis.
\newblock {\em Discourse Processes}, 25(2\&3):285--307, 1998.

\bibitem{FungN06}
P.~Fung and G.~Ngai.
\newblock One story, one flow: Hidden markov story models for multilingual
  multidocument summarization.
\newblock {\em TSLP}, 3(2):1--16, 2006.

\bibitem{GaoNNZWH10}
W.~Gao, C.~Niu, J.~Nie, M.~Zhou, K.~Wong, and H.~Hon.
\newblock Exploiting query logs for cross-lingual query suggestions.
\newblock {\em {ACM} Trans. Inf. Syst.}, 28(2), 2010.

\bibitem{grosz:1983}
B.~J. Grosz, A.~K. Joshi, and S.~Weinstein.
\newblock Providing a unified account of definite noun phrases in discourse.
\newblock In {\em AMACL}, pages 44--50, 1983.

\bibitem{grosz:1986}
B.~J. Grosz and C.~L. Sidner.
\newblock Attention, intentions, and the structure of discourse.
\newblock {\em ACL}, pages 175--204, 1986.

\bibitem{grosz:1995}
B.~J. Grosz, S.~Weinstein, and A.~K. Joshi.
\newblock Centering: A framework for modeling the local coherence of discourse.
\newblock {\em ACL}, pages 203--225, 1995.

\bibitem{Guinaudeau:Strube:2013}
C.~Guinaudeau and M.~Strube.
\newblock Graph-based local coherence modeling.
\newblock In {\em ACL}, pages 93--103, 2013.

\bibitem{HallidayH76}
M.~K. Halliday and R.~Hasan.
\newblock {\em Cohesion in English}.
\newblock Longman, London, 1976.

\bibitem{huffman1952method}
D.~A. Huffman.
\newblock A method for the construction of minimum-redundancy codes.
\newblock {\em IRE}, 40(9):1098--1101, 1952.

\bibitem{joshi:1981}
A.~K. Joshi and S.~Weinstein.
\newblock Control of inference: Role of some aspects of discourse
  structure-centering.
\newblock In {\em IJCAI}, pages 385--387, 1981.

\bibitem{KanungoO09}
T.~Kanungo and D.~Orr.
\newblock Predicting the readability of short web summaries.
\newblock In {\em {WSDM}}, pages 202--211, 2009.

\bibitem{karamanis:2006}
N.~Karamanis.
\newblock Evaluating centering for sentence ordering in two new domains.
\newblock In {\em NAACL}, pages 65--68, 2006.

\bibitem{KaramanisMPO09}
N.~Karamanis, C.~Mellish, M.~Poesio, and J.~Oberlander.
\newblock Evaluating centering for information ordering using corpora.
\newblock {\em ACL}, 35(1):29--46, 2009.

\bibitem{Kehler02}
A.~Kehler.
\newblock {\em Coherence, Reference and the Theory of Grammar}.
\newblock CSLI Publications, California, 2002.

\bibitem{Kibble01}
R.~Kibble.
\newblock A reformulation of rule 2 of centering theor.
\newblock {\em ACL}, 27(4):579--587, 2001.

\bibitem{KibbleP04}
R.~Kibble and R.~Power.
\newblock Optimizing referential coherence in text generation.
\newblock {\em ACL}, 30(4):401--416, 2004.

\bibitem{Lapata03}
M.~Lapata.
\newblock Probabilistic text structuring: Experiments with sentence ordering.
\newblock In {\em ACL}, pages 545--552, 2003.

\bibitem{DBLP:conf/acl/2011}
D.~Lin, Y.~Matsumoto, and R.~Mihalcea, editors.
\newblock {\em ACL 2011}. ACL, 2011.

\bibitem{LinLNK12}
Z.~Lin, C.~Liu, H.~T. Ng, and M.-Y. Kan.
\newblock Combining coherence models and machine translation evaluation metrics
  for summarization evaluation.
\newblock In {\em ACL (1)}, pages 1006--1014. ACL, 2012.

\bibitem{LinNK11}
Z.~Lin, H.~T. Ng, and M.-Y. Kan.
\newblock Automatically evaluating text coherence using discourse relations.
\newblock In Lin et~al. \cite{DBLP:conf/acl/2011}, pages 997--1006.


\balancecolumns



\bibitem{LiomaLL12}
C.~Lioma, B.~Larsen, and W.~Lu.
\newblock Rhetorical relations for information retrieval.
\newblock In {\em {SIGIR}}, pages 931--940, 2012.





\bibitem{louis:2010}
A.~Louis and A.~Nenkova.
\newblock Creating local coherence: An empirical assessment.
\newblock In {\em NAACL}, pages 313--316, 2010.

\bibitem{LouisN12}
A.~Louis and A.~Nenkova.
\newblock A coherence model based on syntactic patterns.
\newblock In {\em EMNLP-CoNLL}, pages 1157--1168. ACL, 2012.

\bibitem{MannT98}
W.~C. Mann and S.~A. Thompson.
\newblock Rhetorical structure theory: Toward a functional theory of text
  organization.
\newblock pages 243--281, 1998.

\bibitem{hinrich}
C.~D. Manning and H.~Sch{\"{u}}tze.
\newblock {\em Foundations of statistical natural language processing}.
\newblock {MIT} Press, 2001.

\bibitem{Mikk:2001}
J.~Mikk.
\newblock Prior knowledge of text content and values of text characteristics.
\newblock {\em Journal of Quantitative Linguistics}, 8(1):67--80, 2001.

\bibitem{MiltsakakiK04}
E.~Miltsakaki and K.~Kukich.
\newblock Evaluation of text coherence for electronic essay scoring systems.
\newblock {\em Natural Language Engineering}, 10(1):25--55, 2004.

\bibitem{NtoulasNMF06}
A.~Ntoulas, M.~Najork, M.~Manasse, and D.~Fetterly.
\newblock Detecting spam web pages through content analysis.
\newblock In L.~Carr, D.~D. Roure, A.~Iyengar, C.~A. Goble, and M.~Dahlin,
  editors, {\em {WWW}}, pages 83--92. {ACM}, 2006.

\bibitem{PitlerN08}
E.~Pitler and A.~Nenkova.
\newblock Revisiting readability: A unified framework for predicting text
  quality.
\newblock In {\em EMNLP}, pages 186--195. ACL, 2008.

\bibitem{PoesioSEH04}
M.~Poesio, R.~Stevenson, B.~D. Eugenio, and J.~Hitzeman.
\newblock Centering: A parametric theory and its instantiations.
\newblock {\em ACL}, 30(3):309--363, 2004.

\bibitem{shannon:2001}
C.~E. Shannon.
\newblock A mathematical theory of communication.
\newblock {\em Bell System Technical Journal}, 3(27):379--423, 1948.

\bibitem{shin:2000}
H.~Shin and J.~F. Stach.
\newblock Using long runs as predictors of semantic coherence in a partial
  document retrieval system.
\newblock In {\em NAACL-ANLP}, pages 6--13, 2000.

\bibitem{sigurd2004word}
B.~Sigurd, M.~Eeg-Olofsson, and J.~Van~Weijer.
\newblock Word length, sentence length and frequency--zipf revisited.
\newblock {\em Studia Linguistica}, 58(1):37--52, 2004.

\bibitem{SoricutM06a}
R.~Soricut and D.~Marcu.
\newblock Discourse generation using utility-trained coherence models.
\newblock In N.~Calzolari, C.~Cardie, and P.~Isabelle, editors, {\em ACL}. ACL,
  2006.

\bibitem{TanGP12}
C.~Tan, E.~Gabrilovich, and B.~Pang.
\newblock To each his own: personalized content selection based on text
  comprehensibility.
\newblock In {\em {WSDM}}, pages 233--242, 2012.

\bibitem{Tapiero07}
I.~Tapiero.
\newblock {\em Situation Models and Levels of Coherence: Towards a Definition
  of Comprehension}.
\newblock Lawrence Erlbaum Associates, Manwah, New Jersey, 2007.

\bibitem{xiong:2013}
D.~Xiong, Y.~Ding, M.~Zhang, and C.~L. Tan.
\newblock Lexical chain based cohesion models for document-level statistical
  machine translation.
\newblock In {\em EMNLP}, pages 1563--1573, 2013.

\bibitem{zhang:2011}
R.~Zhang.
\newblock Sentence ordering driven by local and global coherence for summary
  generation.
\newblock In {\em ACL}, pages 6--11, 2011.

\bibitem{ZhouC05}
Y.~Zhou and W.~B. Croft.
\newblock Document quality models for web ad hoc retrieval.
\newblock In {\em {CIKM}}, pages 331--332, 2005.

\end{small}
\end{thebibliography}
\balancecolumns
\end{document}